\documentclass[12pt,preprint]{aastex}
\usepackage{epsfig}
\usepackage{graphicx}
\usepackage{epstopdf}
\usepackage{amsmath}
\usepackage{amssymb}
\usepackage{appendix}
\def\be{\begin{equation}}
\def\ee{\end{equation}}
\def\ba{\begin{eqnarray}}
\def\ea{\end{eqnarray}}
\def\go{\mathrel{\raise.3ex\hbox{$>$}\mkern-14mu
             \lower0.6ex\hbox{$\sim$}}}
\def\lo{\mathrel{\raise.3ex\hbox{$<$}\mkern-14mu
             \lower0.6ex\hbox{$\sim$}}}

\begin{document}

%\title{Flux Rope Eruptions for Magnetar Giant Flares --- \\
% I. Boundary Effects on Energy Release }
% Eruptions on Energy Release

\title{Twist-induced Magnetosphere Reconfiguration \\
for Intermittent Pulsars }

%\title{ Flux Rope Eruptions for Magnetar Giant Flares --- \\
%I. Boundary Effects on Energy Release }

%\author{draft by Lei Huang\altaffilmark{1}}
\author{Lei Huang\altaffilmark{1}, Cong Yu\altaffilmark{2,3}, Hao Tong\altaffilmark{4}}
\altaffiltext{1}{Key
Laboratory for Research in Galaxies and Cosmology, Shanghai
Astronomical Observatory, Chinese Academy of Sciences, Shanghai,
200030, China; {\tt muduri@shao.ac.cn}}
\altaffiltext{2}{Yunnan Observatories, Chinese Academy of Sciences, Kunming,
650011, China; {\tt cyu@ynao.ac.cn}}

\altaffiltext{3}{Key
Laboratory for the Structure and Evolution of Celestial Object,
Chinese Academy of Sciences, Kunming, 650011, China;
}

\altaffiltext{4}{Xinjiang Astronomical Observatory, Chinese Academy of Sciences,
Urumqi, Xinjiang 830011, China}

\begin{abstract}
We propose that the magnetosphere reconfiguration induced by
magnetic twists in the closed field line region
can account for the mode-switching of intermittent pulsars.
We carefully investigate the properties of axisymmetric
force-free pulsar magnetospheres with magnetic twists in closed
field line region around the polar caps.
The magnetosphere with twisted closed lines leads to enhanced spin-down rates.
The enhancement in spin-down rate depends on the size of region with twisted closed lines.
Typically,  it is increased by a factor of $\sim2$, which is consistent with
the intermittent pulsars' spin down behavior during the `off' and `on' states.
We find there is a threshold of maximal twist angle $\Delta\phi_{\rm thres}\sim1$.
The magnetosphere is stable only if the closed line twist angle is less than $\Delta\phi_{\rm thres}$.
Beyond this value, the magnetosphere becomes unstable and gets untwisted.
The spin-down rate would reduce to its off-state value.
The quasi-periodicity in spin-down rate change can be explained by
long-term activities in star's crust and the untwisting induced by MHD instability.
The estimated duration time of on-state
is about one week, consistent with observations. Due to the MHD instability, there
exists an upper limit for the spin down ratio  ($f\sim3$)
between the on-state and the off-state, if the Y-point remains at the light cylinder.
\end{abstract}


%%%%%%%%%%%%%%%%%%%%%%%%%%%%%%%%%%%%%%%%%%%%% %\citep[][etc.]

\section{Introduction}

Intermittent pulsars are a special kind of transient radio pulsars. They showed
the correlation between timing and radiation of pulsars \citep[e.g.][]{Kramer06, Wang07,Zhang07}.
The first discovered intermittent pulsar, PSR B1931+24, cycled through
the sequence of `on'-states and `off'-states around once a month \citep{Kramer06}.
In the off-states, the pulsar is radio-quiet and has a relatively low spin-down rate.
In the on-states, the radio emission is produced and the spin-down rate
increases by a factor of about $1.5$, i.e., $\dot{\Omega}_{\rm on}/\dot{\Omega}_{\rm off}\sim1.5$.
The on-state and off-state of PSR B1931+24 are observed to recur quasi-periodically.
During each cycle, it keeps in the off-state for $\sim80\%$ of the time.
When it switches to the on-state, it maintains for less time, $\sim20\%$ of one cycle,
i.e., about one week, then switches off again \citep{Kramer06,Young13}.
Another two intermittent pulsars, PSR J1832+0029 \citep{Lorimer12}
, and PSR J1841-0500 \citep{Camilo12} have been discovered.
PSR J1841-0500 shows a even higher spin-down rate ratio between
on- and off-states, $\dot{\Omega}_{\rm on}/\dot{\Omega}_{\rm off}\sim2.5$.
The mode-switching timescale for the latter two intermittent pulsars is on the order of years.

The changes in the spin-down rates may suggest that dramatic reconfiguration
takes place in the intermittent pulsar magnetosphere.
However, the specific physical process in the magnetosphere is still poorly understood.
In general, the global configuration of the pulsar magnetosphere is not pure dipole.
It is believed that the field lines which have foot-points at star's surface
on the polar caps and cross the light cylinder (LC) are opening up.
The last closed line crosses the equatorial plane at a Y-point, which is located exactly at LC
\citep{Contop99}.
The relativistic particles can be accelerated through these open lines.
In previous works, relativistic particles are considered to form currents,
or, winds, which contribute to the radio emission and high spin-down rate
for the on-state of an intermittent pulsar \citep[e.g.][]{XQ01,LiL14,Kou15}.
In its off-state, the particle winds stop
acceleration somehow so that the pulsar shows low luminosity and low spin-down rate.
\citet{LiJ12a,LiJ12b}
modeled the on states using force-free condition in global magnetosphere,
while modeled the off states using vacuum condition only in the open lines region.
Spin-down rate ratios $f\lesssim2.9$ could be obtained with inclination angles
$\gtrsim30^\circ$.
In other work, the mode-switching behavior of intermittent pulsars is thought to be caused
by shifting the Y-point inside LC \citep{Timokhin06,Timokhin10}.

In another kind of neutron stars with ultra-strong magnetic field $B\gtrsim 10^{14}{\rm G}$,
magnetars, the crustal motions are widely believed to correspond to many activities
in observations \citep[]{WT06,Kaspi07,Mere08,Belob09,Yu12,YH13,HY14a,HY14b}.
There is no doubt that the field in magnetars are strong enough to cause crustal motions.
The average field of an ordinary pulsar is only $\sim10^{12}{\rm G}$.
However, observations in X-ray emission from old radio pulsars
imply hot-spots in polar cap region.
Researches on pair creation propose a Partially Screened Gap model
to describe this inner acceleration region \citep[][ref therein]{Szary13,Szary15}.
The hot-spots are found in small scale, compared to the size of polar cap from purely dipolar field.
The local surface magnetic field is estimated as of the order of $10^{14}{\rm G}$,
much stronger than the average field of the pulsar.
Spot-like regions with ultra-strong field are also suggested
in other studies on neutron stars of nearly all ages \citep[e.g.][]{Gepp03,ML06,Storch14}.
Hence, we assume locally ultra-strong field can exist in polar region of an intermittent pulsar,
which is much older than magnetar.
It causes crustal motions and leads naturally to the magnetic twists in the magnetosphere.

In this paper, we propose a twist-induced intermittent pulsar model caused by
local crustal motions around the polar caps.
For simplicity, we consider the neutron star as an aligned rotator.
The crustal motions deform the magnetosphere by twisting a bundle of closed lines
around the polar caps. We quantitatively investigate the effects of the twisted closed lines
on the stationary configuration of magnetosphere.
We interpret the mode-switching of intermittent pulsars
by the twist-induced reconfiguration of magnetosphere.
The surface magnetic field lines are supposed to be highly non-dipolar
in the polar cap region of a pulsar.
Charged particles are accelerated along these curved field lines so that
the radio emission is generated.
However, it is still a puzzling problem that the radio emission is observed to cease
after an intermittent pulsar switches off.
In our model, an on-state magnetosphere with twisted closed lines possesses more open lines,
i.e., larger emission beam, than its off-state.
The generation and nulling of radio emission may be related to the orientation of
the line of sight of observers \citep{Timokhin10}.

The paper is arranged as follows. The basic model is described in Section 2.
In Section 3, we describe two types of magnetospheres representing the on-state and off-state, respectively.
The state transitions are discussed in Section 4.
We further explore the physical constraint on the spin-down rate enhancement of the on-state
magnetospheres in Section 5.
Discussions are provided in Section 6.
Throughout this paper, we set the magnetic dipole moment $\mu$, the pulsar's rotational frequency $\Omega$,
and the speed of light $c$ to unity, i.e., $\mu=\Omega=c=1$.

%%%%%%%%%%%%%%%%%%%%%%%%%%%%%%%%%%%%%%%%%%%%%

\section{Twisting Induced Force-free Pulsar Magnetosphere}

We assume the force-free condition ${\bf j}\times{\bf B} + c\rho_e{\bf E}=0$
is satisfied everywhere in the stationary axisymmetric pulsar magnetosphere.
Here ${\bf j}$ is the electric current density, ${\bf B}$ and ${\bf E}$
are the magnetic field and electric field, and $\rho_e=\nabla\cdot{\bf E}/4\pi$
is the electric charge density in the pulsar magnetosphere.
The force-free condition provides a simple yet accurate approximation to the neutron star's magnetosphere \citep{Contop99,Gruz05}.
We adopt cylindrical coordinates $(R,z,\phi)$, where both $R$ and $z$ are in unit of $R_{\rm LC} \equiv c/\Omega$. For typical rotating frequency of intermittent pulsars $\Omega\sim$10 ${\rm s}^{-1}$,
the value of $R_{\rm LC}$ is about $3\times10^4$ km, measured from the center of star.
The magnetic field of a standard magnetosphere described above reads
\begin{eqnarray}
    {\bf B}&=& -\frac{\Psi_{,z}}{R} \hat{R}\ +\ \frac{\Psi_{,R}}{R} \hat{z}\ +\ \frac{2I}{R} \hat{\phi} \ .
\end{eqnarray}
where $\Psi$ is the magnetic stream function.
In previous work \citep[e.g.][]{Contop99,OK03,Gruz05,Taka14},
the field line configuration of a standard rotating pulsar magnetosphere is
divided into a closed line region and an open line region, separated by a magnetic separatrix
which contains a current sheet on the equatorial plane.
A Y-point structure is formed at the tip of current sheet.
The Y-point is the intersection of the last closed line and the equatorial plane.
The distance from Y-point to the center of star is
equal to the cylindrical radius of LC, $R_{\rm LC}\equiv c/\Omega$.
The electric currents $I(\Psi)$, consisting of relativistic particles,
are flowing along the open field lines.
The Poynting power $L$ can be calculated as \citep{Gruz05,Timokhin06}
\begin{eqnarray}
    L&=& \int_0^{\Psi_{\rm last}} 2|I(\Psi)|\ {\rm d}\Psi\ ,
\end{eqnarray}
where $\Psi_{\rm last}$ is value of $\Psi$ at the last closed field line.
In the force-free limit, the spin-down power,
which describes the radiation originated from the star's rotational energy,
is assumed to be comparable to the Poynting power,
i.e., $L\sim-I_{\rm rot}\Omega\dot{\Omega}$,
where $I_{\rm rot}$ is the neutron star moment of inertia \citep{LK05}.
Therefore, the spin-down rate ratio $f$ between the on-state and off-state of a pulsar,
$f=\dot{\Omega}_{\rm on}/\dot{\Omega}_{\rm off}$, is approximately measured by
$L_{\rm on}/L_{\rm off}$. We also call $L$ the spin-down power in this paper.

In the standard pulsar magnetosphere, magnetic twist exists only in the open line region.
The closed lines region is entirely untwisted. In other words, the electric currents $I$
exist only in a range from $\Psi=0$ to $\Psi=\Psi_{\rm last}$ \citep{Contop99,Gruz05}.
However, ultra-strong field in the region around the polar cap would cause
active crust motion. In these crustal active regions, the foot-points of
the closed lines on star's surface are displaced.
As a result, the magnetic twist is generated by the active crust motion in the closed line region.
We measure the twist by the twist angle
\begin{eqnarray}
\label{TA}
    \Delta\phi(\Psi)&=& 2|I_{\rm closed}(\Psi)|\ \int_{\Psi}\ \frac{d\theta}{r^3\sin^2\theta B_\theta}\ , \Psi_{\rm last}\le\Psi\le \Psi_{\rm m} \ .
\end{eqnarray}
Here $(r,\theta)$ are spherical coordinates and $\Psi_{\rm m}$ is the upper boundary of
twisted closed lines. Basically, $\Psi_{\rm m}$ should be chosen as small values
since the crustal active region cannot be too large.
In practice, the star radius of a pulsar is $R_{\rm NS}\sim10$ km $\sim3\times10^{-4}R_{\rm LC}$.
We choose a typical value for the upper boundary as $\Psi_{\rm m}=2.1$ (see in next section).
Since the magnetosphere near the star's surface is roughly in dipolar pattern \citep{Contop99}, one can find the footpoint on star's surface of this boundary field line
is at $\theta_{\rm m}=\arcsin\sqrt{\Psi_{\rm m}R_{\rm NS}}=1.44^\circ$,
in poloidal direction measured from the star's pole.
Thus, the typical length of the crustal active region on star's surface
can be estimated as $2R_{\rm NS}\sin\theta_{\rm m}\sim0.5{\rm km}$,
which is consistent with
the length scale of ultra-strong field spots proposed for neutron star \citep{Shabaltas12}.
The global configuration of magnetosphere should satisfy the following equation
\begin{eqnarray}
\label{GS}
    &\ & (1-R^2)\left(\Psi_{,RR}+\Psi_{,zz}\right) - \frac{1+R^2}{R} \Psi_{,R} + 4I(\Psi)\frac{{\rm d}I(\Psi)}{{\rm d}\Psi}\ =\ 0.
\end{eqnarray}
Different from the standard magnetosphere \citep{Contop99,Gruz05}, the electric currents $I$ become non-zero
in closed lines region due to active crustal motions around polar caps. The twist triggered by local crustal motion
is expected to induce reconfiguration of the global magnetosphere.

\section{Magnetospheres During On-State and Off-State}

Theoretically, the distribution of poloidal current can be divided into two parts,
$I_{\rm closed}(\Psi_{\rm last} \le \Psi \le \Psi_{\rm m})$ in closed lines region
and $I_{\rm open}(\Psi\le\Psi_{\rm last})$ in open lines region.
While $I_{\rm open}$ can be consistently determined by iteration,
$I_{\rm closed}$ requires a reasonable trial function.
We set a trial of $I_{\rm closed}(\Psi)$ as
\begin{eqnarray}\label{demo}
    I_{\rm closed}(\Psi)&=& \mathcal{A} \ (\Psi-\Psi_{\rm last})(\Psi-\Psi_{\rm m})\ ,\ \Psi_{\rm last}\le\Psi\le\Psi_{\rm m}\ .
\end{eqnarray}
This trial function provides a simple but reasonable description to the current in the closed lines region
induced by crustal motions.
Since we assume stationary magnetosphere during on-state so that the electric
circuit in open lines region is close, which requires $I_{\rm closed}(\Psi_{\rm last})=0$.
We also require that the magnetic twist be locally confined within the polar region
inside $\Psi_{\rm m}$, which indicates that $I_{\rm closed}(\Psi>\Psi_{\rm m})\equiv0$.
The extent of crustal motions can be described by the amplitude $\mathcal{A}$.
The greater the amplitude $\mathcal{A}$, the more twisted the magnetosphere.
Notice that we adopt a simple parabolic profile for $I_{\rm closed}$.
We have tested different current profiles within the closed field line region and
found that, different choices of current profiles indeed bring about some minor quantitative differences,
but the overall qualitative behaviors of the twisted magnetosphere are the
same. For simplicity, we use the current distribution in Equation (\ref{demo}) as a demonstration example for our further discussion.

%numerical details ---

%$I_{\rm open}$ ---  treatment on LC --- singularity
It can be noticed that Eq.\ref{GS} shows property of singularity at LC, where $R=1$.
In order to determine the current distribution in open lines region
$I_{\rm open}(\Psi\le\Psi_{\rm last})$,
we use the L'H\^opital's rule to treat the terms with $\Psi_{,R}$ and $(II')_{\rm open}$
which contain singularities in the equation.
The boundary conditions on LC are taken as
$\left. \Psi_{,R} \right|_{R=1^\pm} = \left. 2(II')_{\rm open} \right|_{R=1^\pm}$.
With an appropriate trial function of $I_{\rm open}$ chosen,
we correct the distribution of $I_{\rm open}$ across LC iteratively with \citep{Contop99}
\begin{eqnarray}
\label{IIp}
    (II')_{\rm open}(\Psi)&=& \mu_1 \left. (II')_{\rm open} \right|_{R=1^+} + \mu_2 \left. (II')_{\rm open} \right|_{R=1^-} + \mu_3 ( \Psi_{R=1^+}-\Psi_{R=1^-} )\ , \nonumber\\
    \Psi&=& \frac{1}{2} \left( \Psi_{R=1^+}+\Psi_{R=1^-} \right)\ ,
\end{eqnarray}
where the weight factors $\mu_1+\mu_2=1$ and $\mu_3$ are chosen empirically.
One more complexity in solving Eq.\ref{GS} is to do with the non-linearity of this equation.
Let us define a linear operator $\mathcal{L}$ to represent the differentiations of $\Psi$,
Eq.\ref{GS} becomes $\mathcal{L}(\Psi)+II'(\Psi)=0$. Obviously, the term $II'$ contains
the complicated non-linear behavior of the equation.
After each correction of $II'$ by Eq.\ref{IIp}, we calculate its differentiation
${\rm d}(II')/{\rm d}\Psi$. We adopt Newton iteration \citep{Press92}
to calculate a new $\Psi$ as
\begin{eqnarray}
    \Psi^{\rm new}&=& \Psi - \frac{\mathcal{L}+II'}{\partial\mathcal{L} /\partial\Psi + {\rm d}(II')/{\rm d}\Psi}\ .
\end{eqnarray}
In practice, it is an effective method in searching the solution accurately.
%non-linearity Newton iteration

%boundary conditions and return current --- see in ref.
In each step of iteration, the poloidal current $I_{\rm open}(\Psi)$
is obtained by integrating $(II')_{\rm open}$.
It should be pointed out that there is a constraint $I_{\rm open}(\Psi_{\rm last})=0$,
which requires the outgoing current return to the star.
In this paper, we assume an ideal current sheet forms on the equatorial plane outside LC,
i.e., $\Psi(R\ge1,0)\equiv \Psi_{\rm last}$.
Then we adopt a return current $I_{\rm ret}$ in a narrow range
$[\Psi_{\rm last}, \Psi_{\rm last}+\delta]$ \citep{Contop99,Gruz05},
to close the electric circuit.
Alternative treatments for close circuit without current sheet assumed
can be found in others' work \citep[e.g.][]{Taka14}.

\begin{figure}
\includegraphics[scale=0.8]{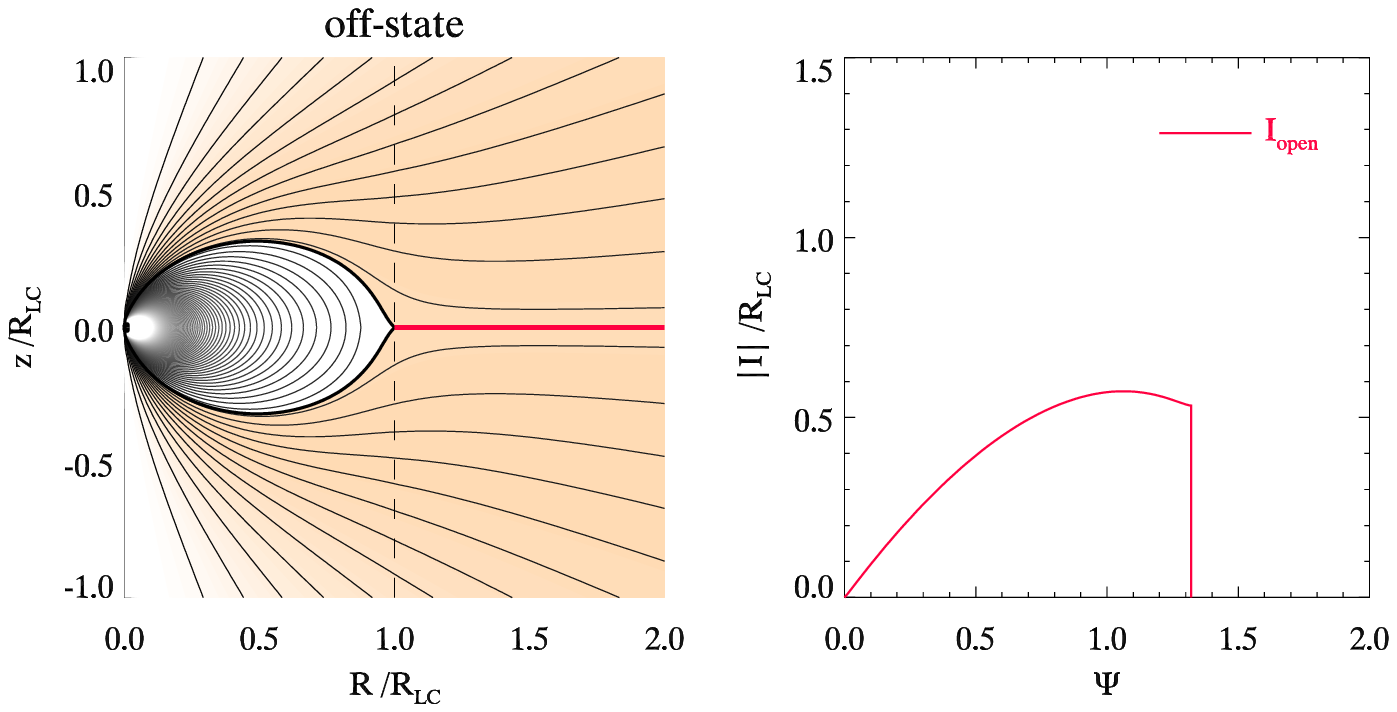} \\
\includegraphics[scale=0.8]{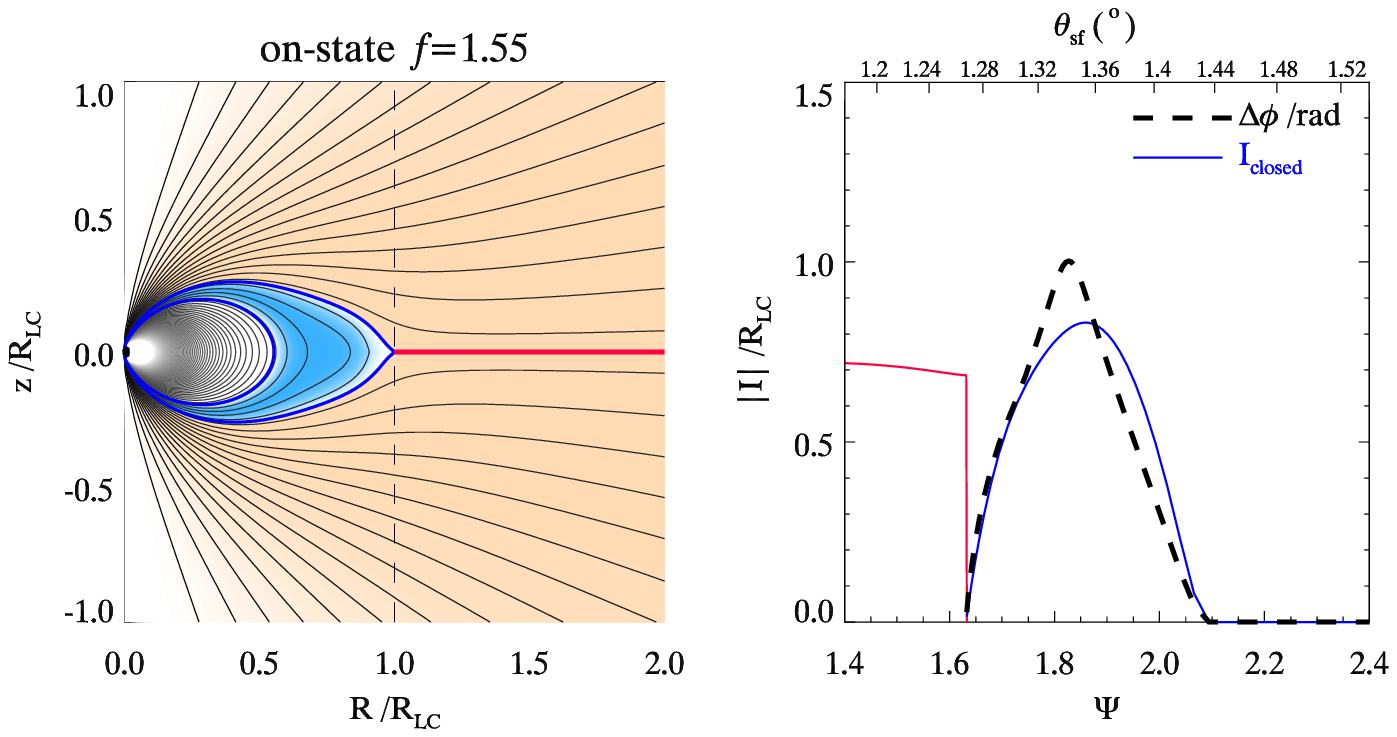}
\caption{\label{twist}
{\it Top-left}: The field lines configuration of pulsar magnetosphere in off-state.
The dashed line represents LC. The thick line represents the magnetic separattrix,
with red part for the current sheet. The red shade represents the currents $|I|$  flowing in open lines.
{\it Top-right}: The distribution of currents in open lines region,
corresponding to the off-state magnetosphere.
{\it Bottom-left}: The field lines configuration of magnetosphere,
as a candidate interpretation of the on-state of PSR B1931+24.
Currents induced by twisting in closed lines are flowing in the blue shaded area.
The spin-down rate ratio $f\approx1.55$.
{\it Bottom-right}: The distribution of $|I|$ in open lines region (solid red) and
closed line region (solid blue). The dashed black line represents the distribution
of corresponding twist angle $\phi(\Psi)$ in closed line region.
The local crustal active region is within the boundary $\theta_{\rm m}=1.44^\circ$.
The maximal twist angle $\Delta\phi_{\rm max}$ reaches the saturated value $\Delta\phi_{\rm thres}\sim1$.
 }
\end{figure}

In Fig.\ref{twist}, we show two kinds of magnetosphere for intermittent pulsars.
The upper panels represent the off-state and the lower panels
represent the on-state.
The field line configuration of the off-state magnetosphere
is shown in the upper-left panel, which indicates a magnetosphere without
active crustal motions. The dashed line denotes the LC.
The thick line designates the magnetic separatrix,
which contains a current sheet on the equatorial plane (marked in red).
The open lines region (red shaded area) and closed lines region (non-shaded area)
are separated by this thick line.
The Y-point structure is located at $(R,z)=(R_{\rm LC},0)$.
The value of $\Psi$ at this last closed line is approximately, $\Psi_{\rm last}\approx1.32$,
which agrees with the standard results \citep{Contop99,Gruz05}. %[][etc.]
Only the open lines region contains the electric currents,
the distribution of which is shown in the solid red line in the upper-right panel.
The spin-down power of the off-state magnetosphere is
$L_{\rm off}\approx1.07\mu^2\Omega^4c^{-3}$.

The lower panels of Fig.\ref{twist} show the field configuration
and the current distribution of the on-state, where  crustal motions give rise
to twisted magnetosphere within the closed field line region around
the polar cap, with the upper boundary in poloidal direction $\theta_{\rm m}=1.44^\circ$,
corresponding to $\Psi_{\rm m}=2.1$.
We show the twisted field configuration caused by local crustal motions in the lower-left panel of Fig.\ref{twist}.
As expected, global reconfiguration of the magnetosphere is induced by magnetic twist in closed line region.
Compared with the off-state magnetosphere, the primary feature of the global reconfiguration is
that more field lines are opening up, i.e., the open line region (red shaded area) is enlarged.
In the on-state, the magnetosphere enters a new state with electric currents flowing
in closed lines (blue shaded area).
The current distribution of the on state are shown in solid lines (red and blue) in the lower-right panel.
The value of $\Psi$ for the last closed line shifts to $\Psi_{\rm last}\approx1.63$.
Correspondingly, the spin-down power is enhanced to $L_{\rm on}\approx1.66\mu^2\Omega^4c^{-3}$.
The ratio of the spin-down power of the on state to the off-state one is approximately, $f\approx1.55$,
which is consistent with the intermittent spin-down behavior of PSR B1931+24.
The steady current distribution in closed lines region is shown in the solid blue line.
The foot-points of the twisting closed lines distributed on star's surface in poloidal direction
within a range of $1.27^\circ\le \theta_{\rm sf} \le1.44^\circ$,
where $\theta_{\rm sf}=\arcsin\sqrt{\Psi R_{\rm NS}}$ . %{\bf How did you calculate this?}.
The distribution of twist angle in closed line region $\Delta\phi$ is shown as the thick dashed line.
The maximum of twist angle $\Delta\phi_{\rm max}$ is about 1.
In the following section, we will see that this on-state magnetosphere
is a critical state which is expected to get untwisted and transit to off-state again.

\section{State Transitions of Twist Induced Magnetospheres}

In our model, the on-state and off-state correspond to two different kind of pulsar magnetospheres
with and without crustal motion induced twist, respectively.
However, its periodicity of mode-switching, or state transition, remains an unresolved issue.
Interestingly, we find that the on-state magnetosphere is in a critical state,
which may lead to the on-off state transition.

\subsection{Critical States of Magnetospheres}
The spin-down rate enhancement of the on-state and off-state magnetosphere mentioned above
is approximately $f\approx1.55$, which is consistent
with the observational spin-down behavior of the intermittent pulsar, PSR B1931+24, during the on and off states.
We confine the crustal motions in a small active region around the polar cap,
within the specific boundary $\theta_{\rm m}=1.44^\circ$. %%%That is, the active region  is
Then we investigate the magnetosphere's response to the different extent of crustal motion,
which leads to various
global magnetospheric configurations with different level of twist in closed line region.

\begin{figure}
\includegraphics[scale=0.9]{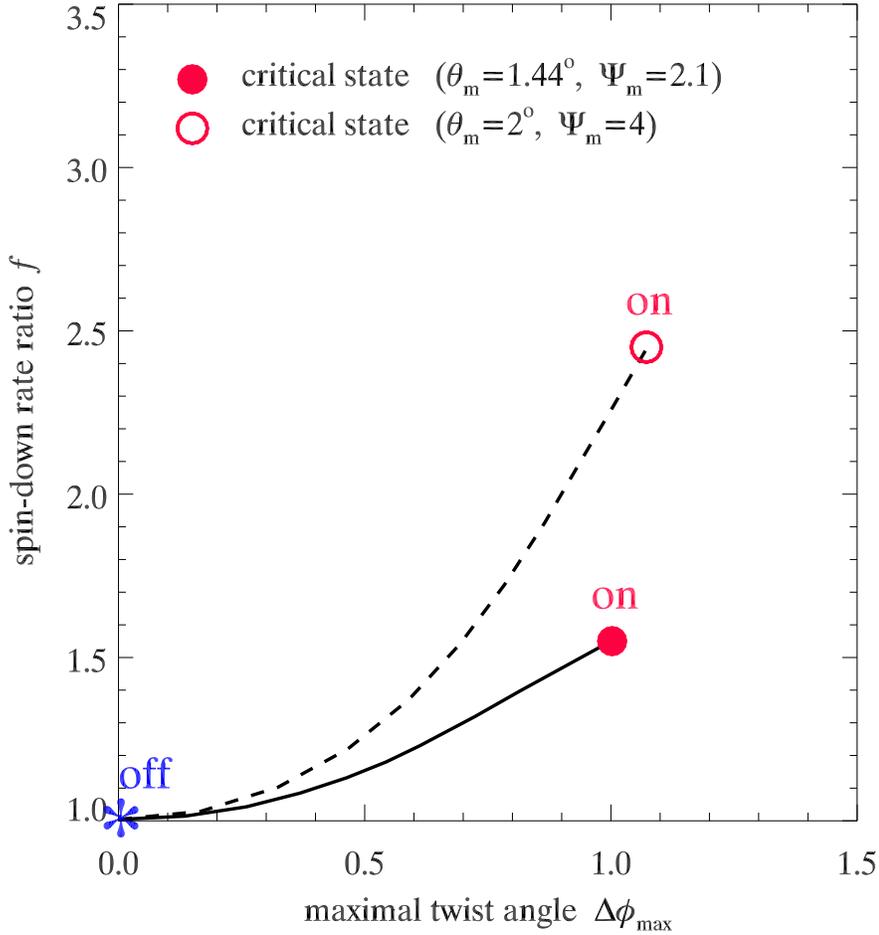}
\caption{\label{lpower}
{\it Solid}: The spin-down rate ratio $f$ in relation to maximal twist angle $\Delta\phi_{\rm max}$,
which implies the twisting level in closed lines region.
The boundary of crustal active region is specified as $\theta_{\rm m}=1.44^\circ$.
The magnetosphere critical state (red filled circle) and the standard magnetosphere (blue star)
can account for the on-state and off-state of PSR B1931+24, respectively.
{\it Dashed}: The same as the solid line but with $\theta_{\rm m}=2^\circ$.
The magnetosphere critical state (red open circle) and the standard magnetosphere
can account for the on-state and off-state of PSR J1841-0500.
 }
\end{figure}

To measure the magnetic twist, we adopt the maximum of the twist angle, $\Delta\phi_{\rm max}$,
in closed line region.
The variation of the spin-down rate ratio $f$ with the maximal twist angle $\Delta\phi_{\rm max}$,
is shown as solid line in Fig.\ref{lpower}.
The left end of the solid line, marked by a star,
represents the standard magnetosphere.
Each point on the curve represents a twisted magnetosphere
in steady state, which indicates that the reconfigured magnetosphere due to a certain level of twist is achievable.
We find that there exists a threshold for the maximum twist angle,
$\Delta\phi_{\rm thres}\approx1$,
marked by a filled circle at the right end of the solid line.
If we increase the twists beyond
this threshold, we are not be able to find equilibrium field configurations any more.
Note that steady state magnetosphere with $\Delta\phi\gtrsim1$
is not achievable because it becomes kink-unstable. Physically, our result is consistent with the MHD instability properties.
The growth of twists would by inhibited by the MHD instability \citep{Uzden02,Belob09}.
The threshold $\Delta\phi_{\rm thres}$ corresponds to a magnetosphere in critical state,
field lines of which would get untwisted and bring the magnetosphere back into a lower energy state.

Higher value of spin-down rate ratio, i.e., $f=2.5$, is observed in PSR J1841-0500.
It can be interpreted by a slightly larger crustal active region.
We adopt $\theta_{\rm m}=2^\circ$, corresponding to $\Psi_{\rm m}=4$.
The results
are shown in the dashed line in Fig.\ref{lpower}.
Similarly, there is a threshold of the maximum of the twist angle, $\Delta\phi_{\rm thres}\approx1$,
marked by an open circle at the right end of the dashed line.

\subsection{State Transitions}

The on-state and off-state of PSR B1931+24 are observed to recur quasi-periodically.
The latter discovered intermittent pulsar, PSR J1832+0029,
shows mode-switching behaviors similar to PSR B1931+24, but with a longer period of $\sim600$ days.
We propose that the periodicity of state transitions
might be due to the long-term active crustal motions around the star's polar cap.
The closed lines there undergo repeated twisting and untwisting, leading to sequent
reconfiguration of the magnetosphere.

The quasi-periodic behavior of the intermittent pulsars, especially for PSR B1931+24, can be naturally
explained within the framework of twist-induced reconfiguration.
As mentioned above, the state with standard magnetosphere and the critical state can account for
the off-state and on-state of pulsar, respectively.
The locally ultra-strong magnetic field stresses the polar crust.
The subsequent crustal motions twist the closed lines significantly.
The reconfiguration induced by the crust motion pushes the magnetosphere to the critical state.
Consequently, more field lines are opened, which leads to  the enhancement in the spin-down rate.
However, the kink instability prevents the magnetosphere to sustain a stable state with further magnetic twists.
The time for the magnetosphere retains the on-state is limited since magnetosphere would get untwisted.
At the end of untwisting process,
the currents in closed lines are dissipated and the magnetosphere
switches back to the off-state, i.e., the standard configuration.
As a result, the spin-down rate decreases. %the radio emission ceases and
With the existence of locally ultra-strong magnetic field,
the polar crusts remain active in the long term.
The pulsar would have periodic behavior of magnetosphere reconfiguration, i.e.,
switch to on-state, due to twist induced by crustal motions,
and switch back to off-state due to untwisting induced by dynamical instabilities.
The spin-down rate during on-states is enhanced by a specific factor,
compared to that during off-states, which may imply that the active region is locally confined.

The generation or nulling of the radio emission is still an unresolved issue.
We interpret the observation of radio emission during on-state with the increasing of
opening angle of the emission beam.
Specifically, the bundle of open lines in the standard magnetosphere form an opening angle
of $\sim1.14^\circ$. This angle increases to $\sim1.27^\circ$ in the
twisted magnetosphere established for PSR B1931+24.
The radiation area is estimated to increase $\sim24\%$.
For PSR J1841-0500, The opening angle increases to $\sim1.4^\circ$ during on-state.
The corresponding radiation area is estimated to increase $\sim50\%$.

The duration time of the on-state of an intermittent pulsar
can be estimated by the untwisting timescale.
By analogy, we consider the untwisting in the polar cap region is similar to the untwisting of global magnetosphere of a magnetar \citep{Belob09}.
The duration of on-state is
$t_{\rm on}\sim 15\ \mathcal{V}_9^{-1} B_{14} R_6^2\ \psi u_*\ {\rm yr}$.
Here, $\mathcal{V}$ is the Voltage in twisted region, $\psi$ is the twist angle,
and $u_*$ is the boundary of region where $|I'(\Psi)|>0$.
We calculate $u_*=\Psi_{\rm m}/\Psi_{\rm R}$ by definition,
where $\Psi_{\rm R}$ is the field line touches the star's surface
at cylindrical coordinates $(R_{\rm NS},0)$.
$\Psi_{\rm R}\sim10^3$ for a typical pulsars' rotating frequency $\Omega\sim10\ {\rm s}^{-1}$.
In model for PSR B1931+24, $\Psi_{\rm m}=2.1$ so that $u_*\sim10^{-3}$.
We adopt $\mathcal{V}=10^9\ {\rm V}$ and $B=10^{14}\ {\rm G}$,
since the field environment of crustal active polar cap of an intermittent pulsar
is expected to be comparable to magnetar.
$\psi$ is assumed to be equal to $\Delta\phi_{\rm thres}\approx1$.
The estimate of $t_{\rm on}$ is $\sim$ 6 days.
This is consistent with duration time of about one week of the on-state in observations.
Another two intermittent pulsars stay in on-states for months to one year.
If the polar region possess locally ultra-strong magnetic field, $B\sim10^{15}\ {\rm G}$,
the relevant timescale could provide appropriate interpretation.
It should be noted that the timescale we estimate is from the pulsar's switch-on to its switch-off.
The spin-down rate is not expected to decrease
immediately after the pulsar switches on. The on-state magnetosphere would
keep the high spin-down rate for a certain period.
Time-dependent calculations are required to describe the details in the state transitions, which
will be reported elsewhere.

\section{Spin-down Rate Enhancement of On-State Magnetospheres}

\begin{figure}
\includegraphics[scale=1]{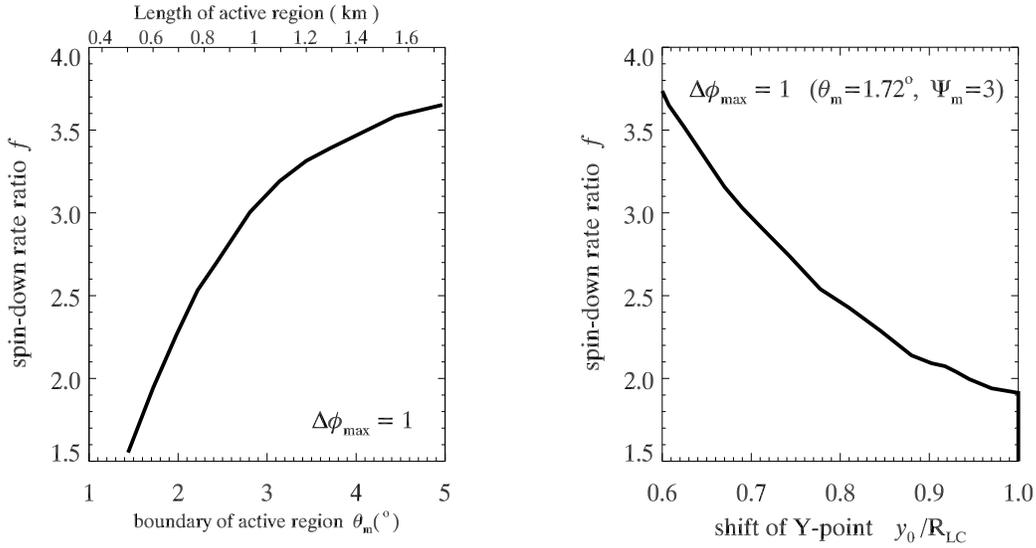}
\caption{\label{y0}
{\it Left}: The spin-down rate ratio $f$ in relation to the boundary of crustal active region
$\theta_{\rm m}$. The maximal twist angle in closed lines region
is specified as $\Delta\phi_{\rm max}=1$.
The corresponding length scale of the active region is shown in the upper abscissa.
The value of $f$ is not expected to be higher than 3, if assuming the length scale is less than 1km.
{\it Right}: The spin-down rate ratio $f$ in relation to the shift of Y-point.
The boundary of crustal active region is specified as $\theta_{\rm m}=1.72^\circ$.
The value of $f$ higher than observations (2.5)
can be obtained with $y_0\lesssim0.8R_{\rm LC}$.
 }
\end{figure}

The three discovered intermittent pulsars have spin-down rate ratios $f\lesssim2.5$.
It leaves an interesting question whether the spin-down rate ratio could be higher.
Theoretically, there exists an upper limit for the spin-down rate enhancement of the on-state magnetosphere since
the active region is locally confined and the closed lines cannot
twist beyond $\Delta\phi_{\rm thres}$.
In the left panel of Fig.\ref{y0}, we show theoretical expectation of
the spin-down rate ratio $f$ in relation to
the size of the crustal active polar cap.
The lower abscissa shows $\theta_{\rm m}$, the boundary of active region
in poloidal direction measured from star's pole.
The upper abscissa shows the corresponding length of active region estimated as
$2R_{\rm NS}\sin\theta_{\rm m}$.
Obviously, a pulsar with larger size of crustal active region is expected to have
higher spin-down rate enhancement in its on-state.
For example, a size $\sim0.5{\rm km}$ corresponds to $f\approx1.5$,
and a size $\sim0.8{\rm km}$ corresponds to $f\approx2.5$.
The size of crustal active region may be less than $1{\rm km}$, i.e.,
the crust thickness of the neutron star \citep{LP04}.
One can find that high spin-down rate ratios $f\gtrsim3$ are not expected
due to the MHD instability.

However, it should be noted that our model assumes the location of Y-point
is fixed at LC when the magnetosphere switches to the on-state.
If the Y-point could be shifted inside LC during the on-state,
higher values of $f$ can be achieved.
The investigation of states of pulsars purely affected by Y-point shifts
can be referred in \citet{Timokhin06}.
Numerically, drawing the location of Y-point is achieved by changing the boundary condition
on the equatorial plane with $\Psi(R\ge y_0/R_{\rm LC},0)\equiv \Psi_{\rm last}$,
where $y_0(<R_{\rm LC})$ represents the current sheet tip extended inside LC.
The rest of numerical treatments is similar to those
in searching solutions with Y-points at LC.
We specify the crustal active region with $\theta_{\rm m}=1.72^\circ$,
i.e., length of $\sim0.65{\rm km}$, and $\Delta\phi_{\rm max}=1$.
In the right panel of Fig.\ref{y0}, we show theoretical expectation of the spin-down rate ratio $f$
in relation to the shift of Y-point, $y_0$ in $R_{\rm LC}$.
Spin-down rate ratios higher than current observations, $f\gtrsim2.5$ can be obtained
with $y_0\lesssim0.8R_{\rm LC}$.
Even higher spin-down rate ratios $f\gtrsim3$ can be obtained with $y_0\lesssim0.7R_{\rm LC}$.

\section{Discussions}

The physics of the unique phenomena of intermittent pulsars is still poorly understood.
We propose a simple model to interpret the periodic mode-switching of intermittent pulsars.
The pulsars possess modest magnetic field $\sim10^{12}{\rm G}$ in average.
However, ultra-strong field $\gtrsim10^{14}{\rm G}$ may exist on star's polar caps within spatially confined region \citep{Gepp03}.
The polar crusts of star are active in the long term caused by ultra-strong field.
The crustal motions trigger magnetic twists in the closed field lines.
Global reconfiguration in the pulsar magnetosphere is induced so that the pulsar
switches to on-state. More field lines are open, resulting in an enhancement in spin-down rate.
The maximal twist angle in the closed lines region of the on-state magnetosphere
reaches a threshold $\Delta\phi_{\rm thres}\sim 1$.
The on-state magnetosphere stays in a critical steady state.
Twisting in closed lines region beyond $\Delta\phi_{\rm thres}$ would induce instability,
so that the magnetosphere would get untwisted and reconfigure to off-state pattern.
As a result, the spin-down rate decreases.
The long-term crustal motions continually trigger magnetic twists.
Hence, the magnetosphere would undergo periodic off-on and on-off reconfiguration
due to twisting and untwisting in the closed lines region.
As long as the crustal active region is finite, the spin-down rate ratio of on-state and off-state
magnetospheres, $f=\dot{\Omega}_{\rm on}/\dot{\Omega}_{\rm off}$
is a specific value during each period of mode-switching.
Crustal active region on star's surface with boundary in poloidal direction
$\theta_{\rm m}=2^\circ$ corresponds to $f\approx2.5$.
This value is adequate to interpret the spin-down rate enhancement
of all discovered intermittent pulsars.
This model may also interpret the spin-down rate change, $f\approx1.36$,
observed in PSR B0540-69 \citep{Marsh15}.

The model proposed in this paper is based on stationary solutions of axisymmetric magnetosphere.
It focuses on the origin of mode-switching of
the intermittent pulsars. The off-state of a pulsar is interpreted by a standard magnetosphere
\citep{Contop99,Gruz05}.
The on-state is interpreted by a critical steady magnetosphere with threshold twisting
in closed lines region, i.e., $\Delta\phi_{\rm max}=\Delta\phi_{\rm thres}$.
Considering the limitation of stationary solutions, the detailed temporal process of state transitions is beyond the scope of this paper.
However, by analogy with the untwisting process in magnetar \citep{Belob09},
we can roughly estimate the timescale of duration when the magnetosphere is in its on-state.

Recent work provides a new type of magnetosphere with different boundary condition
on the current sheet \citep{Contop14}. We also consider twisting induced
reconfiguration based on this new boundary treatment on the current sheet.
With the boundary of crustal active region, $\theta_{\rm m}=1.72^\circ$,
the magnetosphere reaches its critical state with $\Delta\phi_{\rm sat}\sim0.72$.
The spin-down rate ratio $f\approx1.57$ and the corresponding untwisting timescale is 8 days.
The results are consistent with the model based on the standard magnetosphere described above.
Additionally, the results provide good interpretation for observations of PSR B1931+24.

We further study the theoretical constraint of the spin-down rate ratio $f$.
We find that the value of $f$ is confined by length scale of the crustal active region
and the MHD kink instability. If the length scale is comparable to the crust thickness,
the MHD kink instability would prevent high enhancement in spin-down with $f\gtrsim3$.
However, theoretical values of $f\gtrsim3$ can be obtained if the Y-point shifts inside LC
during on-off transition induced by untwisting process.
The detailed investigation on state transitions would be achieved by time-dependent calculation
in future work. Moreover, consideration of physics which determine the twist angle distribution
in closed lines region would improve this model in future work.

\newpage

%%%%%%%%%%%%%%%%%%%%%%%%%%%%%%%%%%%
\acknowledgments  LH and CY thank Dr. Zhoujian Cao for discussions. LH thanks the support by the National Natural
Science Foundation of China (Grants 11203055). CY is grateful for financial support
by the National Natural Science Foundation of China (grants 11373064, 11521303), Yunnan Natural
Science Foundation (grant 2012FB187, 2014HB048) and the Youth Innovation Promotion Association, Chinese Academy of Sciences.
Part of the computation is
performed at HPC Center, Yunnan Observatories, CAS, China. %super-computer in AMSS.

\clearpage

\newpage

%%%%%%%%%%%%%%%%%%%%%%


\begin{thebibliography}{99}

    \bibitem[\protect\citeauthoryear{Beloborodov}{2009}]{Belob09}
        Beloborodov, A. M. 2009, ApJ, 703, 1044

    \bibitem[\protect\citeauthoryear{Camilo et al.}{2012}]{Camilo12}
        Camilo, F., Ransom, S. M., Chatterjee, S., Johnston, S., \& Demorest, P.
2012, ApJ, 746, 63

    \bibitem[\protect\citeauthoryear{Contopoulos, Kazanas, \& Fendt}{Contopoulos et al.}{1999}]{Contop99}
        Contopoulos, I., Kazanas, D., \& Fendt, C. 1999, ApJ, 511, 351

    \bibitem[\protect\citeauthoryear{Contopoulos, Kalapotharakos, \& Kazanas}{Contopoulos et al.}{2014}]{Contop14}
        Contopoulos, I., Kalapotharakos, C., \& Kazanas, D. 2014, ApJ, 781, 46

    \bibitem[\protect\citeauthoryear{Geppert, Rheinhardt, \& Gil}{Geppert et al.}{2003}]{Gepp03}
        Geppert, U., Rheinhardt, M., \& Gil, J. 2003, A\&A, 412, L33

    \bibitem[\protect\citeauthoryear{Gruzinov}{2005}]{Gruz05}
        Gruzinov, A. 2005, PhRvL, 94, 021101

    \bibitem[\protect\citeauthoryear{Huang \& Yu}{2014a}]{HY14a}
        Huang, L. \& Yu, C. 2014, ApJ, 784, 168

    \bibitem[\protect\citeauthoryear{Huang \& Yu}{2014b}]{HY14b}
        Huang, L. \& Yu, C. 2014, ApJ, 796, 3

    \bibitem[\protect\citeauthoryear{Kaspi}{2007}]{Kaspi07}
        Kaspi, V. M. 2007, Ap\&SS, 308, 1

    \bibitem[\protect\citeauthoryear{Klimchuk \& Sturrock}{1989}]{KS89}
        Klimchuk, J. A. \& Sturrock, P. A. 1989, ApJ, 345, 1034

    \bibitem[\protect\citeauthoryear{Kou, Ou, \& Tong}{Kou et al.}{2015}]{Kou15}
        Kou, F. F., Ou, Z. W., \& Tong, H. 2015, arxiv: astro-ph/1507.00643

    \bibitem[\protect\citeauthoryear{Kramer et al.}{2006}]{Kramer06}
        Kramer, M., Lyne, A. G., O¡¯Brien, J. T., Jordan, C. A., \& Lorimer, D. R.
        2006, Science, 312, 549

    \bibitem[\protect\citeauthoryear{Lattimer \& Prakash}{2004}]{LP04}
        Lattimer, J. M. \& Prakash, M. 2004, Science, 304, 536

    \bibitem[\protect\citeauthoryear{Li, Spitkovsky, \& Tchekhovskoy}{Li et al.}{2012a}]{LiJ12a}
        Li, J., Spitkovsky, A., \& Tchekhovskoy, A. 2012, ApJ, 746, 60

    \bibitem[\protect\citeauthoryear{Li, Spitkovsky, \& Tchekhovskoy}{Li et al.}{2012b}]{LiJ12b}
        Li, J., Spitkovsky, A., \& Tchekhovskoy, A. 2012, ApJ, 746, L24

    \bibitem[\protect\citeauthoryear{Li et al.}{2014}]{LiL14}
        Li, L., Tong, H., Yan, W. M., Yuan, J. P., Xu, R. X., \& Wang, N.
        2014, ApJ, 788, 16

    \bibitem[\protect\citeauthoryear{Lorimer \& Kramer}{2005}]{LK05}
        Lorimer, D. R., \& Kramer, M. 2005, Handbook of Pulsar Astronomy (Cambridge:
Cambridge Univ. Press)

    \bibitem[\protect\citeauthoryear{Lorimer et al.}{2012}]{Lorimer12}
        Lorimer, D. R., Lyne, A. G., McLaughlin, M. A., Kramer, M., Pavlov, G. G., \& Chang, C. 2012, ApJ, 758, 141

    \bibitem[\protect\citeauthoryear{Marshall et al.}{2015}]{Marsh15}
        Marshall, F. E., Guillemot, L., Harding, A. K., Martin, P., \& Smith, D. A. 2015, ApJ, 807, L27

    \bibitem[\protect\citeauthoryear{Medin \& Lai}{2006}]{ML06}
        Medin, Z. \& Lai, D. 2006, PhRvA, 74, 062508

    \bibitem[\protect\citeauthoryear{Mereghetti}{2008}]{Mere08}
        Mereghetti, S. 2008, A\&AR, 15, 225

    \bibitem[\protect\citeauthoryear{Ogura \& Kojima}{2003}]{OK03}
        Ogura, J. \& Kojima, Y. 2003, PThPh, 109, 619

    \bibitem[\protect\citeauthoryear{Press et al.}{1992}]{Press92}
        Press, W. H., Teukolsky, S. A., Vetterling, W. T., \& Flannery, B. P. 1992,
Numerical Recipes in C (2nd ed.; Cambridge: Cambridge Univ. Press)

    \bibitem[\protect\citeauthoryear{Shabaltas \& Lai}{2012}]{Shabaltas12}
        Shabaltas, N. \& Lai, D. 2012, ApJ, 748, 148

    \bibitem[\protect\citeauthoryear{Storch et al.}{2014}]{Storch14}
        Storch, N. I., Ho, W. C. G., Lai, D., Bogdanov, S., \& Heinke, C. O. 2014, ApJ, 789, L27

    \bibitem[\protect\citeauthoryear{Szary, Melikidze, \& Gil}{2013}]{Szary13}
        Szary, A., Melikidze, G. I., \& Gil, J. 2013, ASPC, 466, 125

    \bibitem[\protect\citeauthoryear{Szary, Melikidze, \& Gil}{2015}]{Szary15}
        Szary, A., Melikidze, G. I., \& Gil, J. 2015, MNRAS, 447, 2295

    \bibitem[\protect\citeauthoryear{Takamori et al.}{2014}]{Taka14}
        Takamori, Y., Okawa, H., Takamoto, M, \& Suwa, Y. 2014, PASJ, 66, 25

    \bibitem[\protect\citeauthoryear{Timokhin}{2006}]{Timokhin06}
        Timokhin, A. N. 2006, MNRAS, 368, 1055

    \bibitem[\protect\citeauthoryear{Timokhin}{2010}]{Timokhin10}
        Timokhin, A. N. 2010, MNRAS, 408, L41

    \bibitem[\protect\citeauthoryear{Uzdensky}{2002}]{Uzden02}
        Uzdensky, D. A. 2002, ApJ, 574, 1011

    \bibitem[\protect\citeauthoryear{Wang, Manchester, \& Johnston}{Wang et al.}{2007}]{Wang07}
        Wang, N., Manchester, R. N., \& Johnston, S. 2007, MNRAS, 377, 1383

    \bibitem[\protect\citeauthoryear{Wood \& Thompson}{2006}]{WT06}
        Woods, P. M., \& Thompson, C. 2006, in Compact Stellar X-Ray Sources, ed.
        W. H. G. Lewin \& M. van der Klis (Cambridge: Cambridge Univ. Press),
        547

    \bibitem[\protect\citeauthoryear{Xu \& Qiao}{2001}]{XQ01}
        Xu, R. X. \& Qiao, G. J. 2001, ApJ, 561, L85

    \bibitem[\protect\citeauthoryear{Young et al.}{2013}]{Young13}
        Young, N. J., Stappers, B. W., Lyne, A. G., Weltevrede, P., Kramer, M., \& Cognard, I. 2013, MNRAS, 429, 2569

    \bibitem[\protect\citeauthoryear{Yu}{2012}]{Yu12}
        Yu, C., 2012, ApJ, 757, 67

    \bibitem[\protect\citeauthoryear{Yu \& Huang}{2013}]{YH13}
        Yu, C., \& Huang, L., 2013, ApJ, 771, L46

    \bibitem[\protect\citeauthoryear{Zhang, Gil, \& Dyks}{Zhang et al.}{2007}]{Zhang07}
        Zhang, B., Gil, J., \& Dyks, J. 2007, MNRAS, 374, 1103





\end{thebibliography}
\end{document}